\documentclass[journal,comsoc]{IEEEtran}

\usepackage[T1]{fontenc}
\usepackage[cmintegrals]{newtxmath}
\usepackage{tabularx}
\usepackage{cite}
\usepackage{amsmath,amsfonts}
\usepackage{color}
\usepackage{graphicx}
\usepackage{adjustbox}
\usepackage{tabularx}

\usepackage{algpseudocode,algorithm,algorithmicx}
\usepackage{kotex}
\usepackage[center]{caption}
\usepackage{comment}

\newtheorem{theorem}{Theorem}

\newtheorem{lemma}{Lemma}

\begin{document}
\title{SWIPT-enabled NOMA in Distributed Antenna System with Imperfect Channel State Information for Max-Sum-Rate and Max-Min Fairness}
\author{Dongjae Kim,~\IEEEmembership{Member,~IEEE},~Minseok Choi,~\IEEEmembership{Member,~IEEE}
        and~Dong-Wook Seo,~\IEEEmembership{Senior Member,~IEEE}
\thanks{Dongjae Kim is with the Artificial Intelligence Convergence Research Center for Regional Innovation, Korea Maritime \& Ocean University, Busan 49112, South Korea (e-mail: kdj6306@gmail.com).}
\thanks{Minseok Choi is with the Department of Electronic Engineering, Kyung Hee University, Yongin 17104, South Korea (e-mail: choims@khu.ac.kr)}
\thanks{Dong-Wook Seo is with the Department of Radio Communication Engineering, Korea Maritime \& Ocean University, Busan 49112, South Korea (e-mail: dwseo@kmou.ac.kr)}
\thanks{Corresponding authors: Minseok Choi and Dong-Wook Seo}\vspace{-5mm}}

\markboth{}%
{Shell \MakeLowercase{\textit{et al.}}: Bare Demo of IEEEtran.cls for IEEE Communications Society Journals}
\maketitle
    \begin{abstract}
    Motivated by the fact that the data rate of non-orthogonal multiple access (NOMA) can be greatly increased with the help of the distributed antenna system (DAS), we presents a framework in which the DAS contributes not only to the data rate but also the energy harvesting of simultaneous wireless information and power transfer (SWIPT) enabled NOMA.
    This study considers the sum-rate maximization problem and the max-min fairness problem for SWIPT-enabled NOMA in DAS and proposes two different schemes of power splitting and power allocation for SWIPT and NOMA, respectively, with imperfect channel state information (CSI). 
    Numerical results validate the theoretical findings and demonstrate that the proposed framework of using SWIPT-enabled NOMA in DAS achieves the higher data rates than the existing SWIPT-enabled NOMA while guaranteeing the minimum harvested energy.
    \end{abstract}


\IEEEpeerreviewmaketitle
\vspace{-2mm}
    \section{Introduction}
    \IEEEPARstart{T}{he} exponential growth of Internet of Things (IoT) technologies with battery-limited and connected devices has triggered a dramatic increase of energy consumption \cite{CST2015AlFuqaha}.
    Furthermore, replacing wireless device batteries or recharging via wired connections requires excessive operational costs in several IoT scenarios, for example, wireless medical sensors. 
    Fortunately, these IoT devices and sensors generally consume very small power; therefore, the lifetime of IoT devices can be extended by employing wireless power transfer (WPT) \cite{ISIT2008Varshney}. 
    In addition, because IoT devices should repeatedly communicate with their controllers for monitoring and status reporting, simultaneous wireless information and power transfer (SWIPT) has emerged in which the same RF signal can be used to transmit information and energy simultaneously \cite{CommMag2014Krikidis}.
    
    In parallel, non-orthogonal multiple access (NOMA) has been explored in depth to efficiently handle ever-increasing data traffic and mobile devices \cite{ComMag2017Ding}. 
    In particular, the power-domain NOMA improves the spectral efficiency by supporting multiple users in the same frequency band with different power levels, relying on successive interference cancellation (SIC). 
    The integration of SWIPT and NOMA has been recently studied to achieve improved spectral efficiency and energy efficiency \cite{JSAC2016Liu,JSAC2019Tang}.
    
    However, SWIPT-enabled NOMA has a limitation in that the energy transfer efficiency and spectral efficiency rapidly decrease according to the transmission distance.
    Here, the distributed antenna system (DAS) could be a solution to the distance-induced innate problems of SWIPT and NOMA. 
    In DAS, multiple remote radio units (RRUs) are geographically distributed in a given cell and cover subsets of the entire cellular region.
    The cooperation of RRUs and the base station (BS) can increase the system throughput \cite{TWC2007Choi} or reduce the transmit power consumption \cite{TCOM2015Kim}.
    As demonstrated in \cite{IoT2018Huang}, the energy efficiency can be improved by applying SWIPT to DAS.
    In addition, the framework of using NOMA in DAS was proposed, and the synergistic effects of NOMA and DAS were observed in \cite{Access2021Kim}. 
    However, the application of NOMA-enabled SWIPT to DAS has not yet been investigated in depth. 
    
    The contributions of this paper are as follows:
    \begin{itemize}
        \item This paper proposes a framework for applying DAS to the SWIPT-enabled NOMA system and analyze impacts of cooperation with remote radio units (RRUs) of the DAS on the SWIPT-NOMA system.
        In particular, RRUs of the DAS can assist energy harvesting (EH) and information decoding (ID) for users with weak channel conditions from the central unit (i.e., IoT controller) to improve the system throughput.
      
        \item We derive the closed-form expressions of data rates of SWIPT-enabled NOMA in DAS, when estimation of CSI is not accurate.
        In addition, we jointly optimize power splitting (PS) ratios for SWIPT and power allocation for NOMA signaling in the DAS in two problem settings: 1) sum-rate maximization with a minimum data rate constraint and 2) max-min fairness problem.
		
		\item The simulation results demonstrate that SWIPT-enabled NOMA in DAS with the optimal PS ratios and power allocation can boost the system throughput more than SWIPT-OMA in DAS or SWIPT-NOMA only, while guaranteeing the minimum harvested energy. 
		The effect of the level of channel uncertainties was also observed. 
    \end{itemize}
   
    \section{System Model}
    This section introduces the architecture of the DAS and SWIPT-enabled NOMA system model with DAS.
    
    \subsection{Architecture of Distributed Antenna System}
    \label{subsec:system_das}
\begin{figure}[t]
    	\centering
    	\includegraphics[width=0.45\textwidth, trim=-0.5cm 0 0 0]{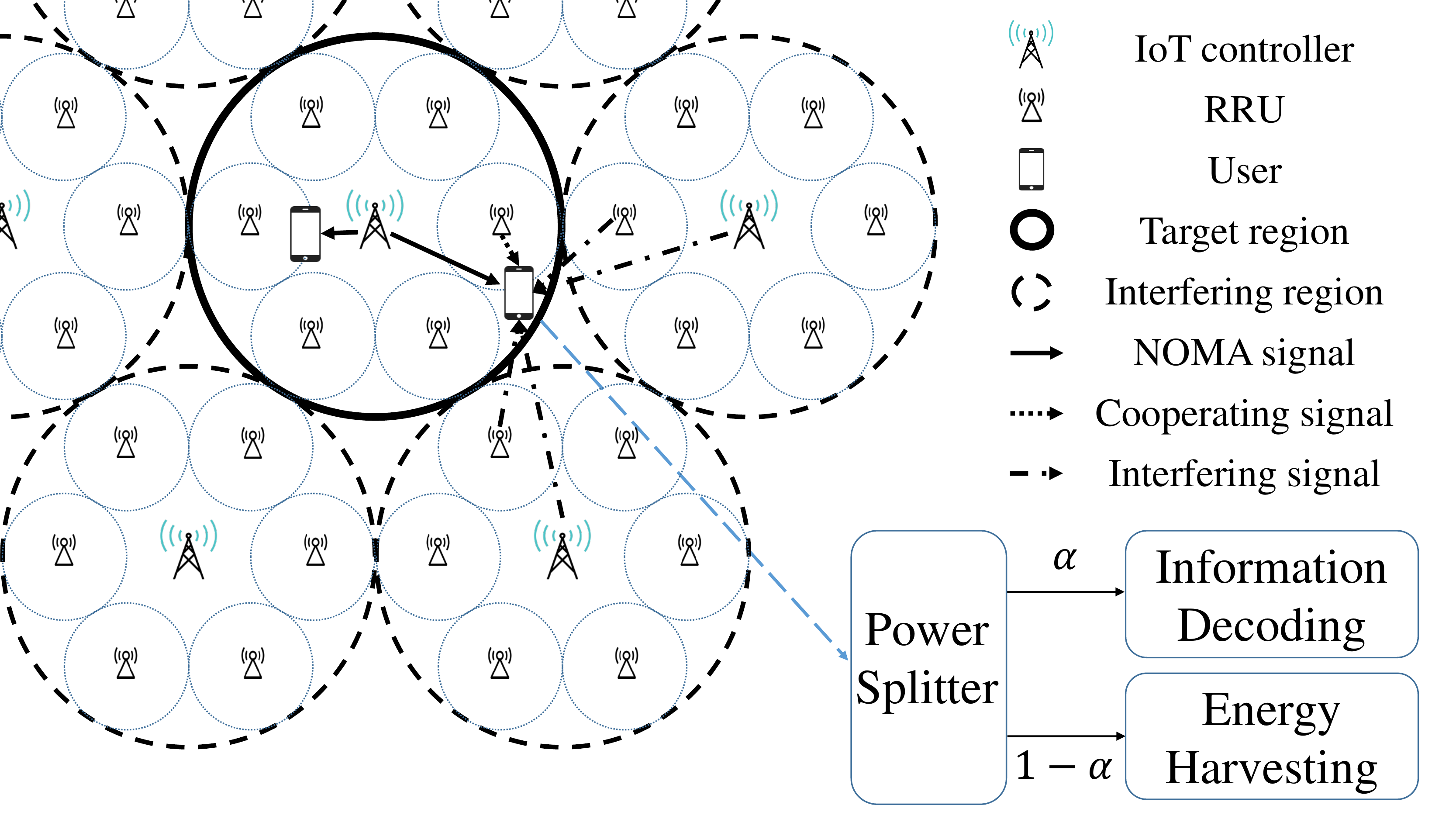}
    	\caption{System Model for SWIPT-enabled NOMA in DAS} \label{fig:system_model}
    	\vspace{-2mm}
    \end{figure}

    We introduce the concept of a general multicell DAS \cite{TWC2007Choi} to the IoT scenario operating SWIPT-enabled NOMA, as shown in Fig. \ref{fig:system_model}.
    A target region consisting of an IoT controller and multiple IoT devices includes $S$ RRUs. 
    
    As in DAS, the IoT controller can support all devices in its target region, while RRUs support devices in their own local areas only.  
    Note that Fig. \ref{fig:system_model} shows an example scenario with $S=6$ RRUs.
    Let $P_m$ and $P_{r}$ denote the power budgets of the IoT controller and RRU, respectively. 
    Then, the total power budget within a region becomes $P = P_m+SP_{r}$.
    Moreover, we assume that there are $L$ dominant interfering regions around the target region.
    
    Within each region, $N$ users are divided into $K$ clusters for NOMA signaling, and $M$ users are grouped for each cluster, that is, $N = MK$.
    We assume that the total bandwidth $B$ is divided into $K$ subbands and users in each cluster are served by using orthogonal subbands.
    The 3GPP LTE Advanced \cite{3GPP} adopts the pairing of two users (or four users optionally) for NOMA signaling because the complexity and error propagation of SIC increases as the number of users sharing the identical band grows. 
    Accordingly, this study assumes a two-user NOMA, that is, $M=2$ and $N=2K$. 
    For simplicity of analysis, we consider a single cluster (i.e., two users) in each region, but our analysis can be easily extended to a general multicluster scenario.
	
	This study is based on a single selection scheme for DAS \cite{TWC2007Choi}, in which each user is served by a central unit (i.e., BS and IoT controller) or only one RRU with the strongest channel condition.
	However, in our system model, the IoT controller pairs a user in the center of the target region (i.e., not in any local region of the RRUs) and a user in the local region of one of the RRUs for NOMA signaling, and they are in the same cluster. 
	As in the single selection scheme \cite{TWC2007Choi}, RRUs also support users in their local regions.
	Here, for simplicity of denotation, the user in the local region of the RRU is called a weak user or user 1, and the user in the center of the target region is called a strong user or user 2. 
	\vspace{-1mm}
    \subsection{Channel Model}
	
	The Rayleigh fading channel from RRU $i$ in region $k$ to user $j$ in the target region is denoted by $h_{i,j}^{(k)}=\sqrt{L_{i,j}^{(k)}} g_{i,j}^{(k)}$, where $L_{i,j}^{(k)}$ and $g_{i,j}^{(k)}\sim CN(0,1)$ are the path-loss and fast fading component, respectively, for all $i\in \lbrace 0, 1, \cdots, S \rbrace$, $j\in \lbrace 1, 2 \rbrace$, and $k\in \lbrace 0, 1, \cdots \rbrace$. 
	The IoT controller is indexed as $i=0$, and the target region is indexed as $k=0$. 
	Here, $L_{i,j}^{(k)}=1/[d_{i,j}^{(k)}]^{\beta}$, where $d_{i,j}^{(k)}$ is the distance between RRU $i$ in region $k$ to user $j$ in the target region, and $\beta$ is the path-loss exponent.
	For simplicity of notation, we drop the region index $k$ of the target region, that is, $h_{i,j}=h_{i,j}^{(0)}$, $L_{i,j}=L_{i,j}^{(0)}$, and $g_{i,j}=g_{i,j}^{(0)}$.
	
	We assume imperfect CSI, and $\hat{h}_{i,j}$ denote an estimate of $h_{i,j}$.
    Suppose that the slow fading component $L_{i,j}$ is known at the transmitter, but the estimation of the fast fading component $g_{i,j}$ is not accurate.
	According to \cite{VTC2019Zamani}, $\hat{h}_{i,j}$ and estimation error $z_{i,j}$ are modeled by
	\begin{equation}\label{eq:channel model}
	    h_{i,j}=\hat{h}_{i,j}+z_{i,j}=\sqrt{L_{i,j}}(\hat{g}_{i,j}+\epsilon_{i,j}),
	\end{equation}
	where $\epsilon_{i,j}\sim CN(0,\sigma_\epsilon^2)$ is the estimation error of $g_{i,j}$ and $\hat{g}_{i,j}$ is the estimated fast fading gain that is uncorrelated with $\epsilon_{i,j}$.

	\subsection{SWIPT-enabled NOMA in DAS}\label{subsec:swipt noma}
	
	As explained in Section \ref{subsec:system_das}, the IoT controller transmits the NOMA signal to both strong and weak users in the same cluster, and RRU $q$ has the strongest channel condition to the weak user, that is, $q=\text{arg}\max_{p\in \lbrace 1, 2, \cdots , S \rbrace}|\hat{h}_{p,1}|^2$, also sends the cooperation signal to the weak user (i.e., user 1). 
	Denote $x_{i,j}$ as the data symbol transmitted from RRU $i$ to user $j$ for all $i\in \lbrace 0, 1, \cdots , S \rbrace$, and $j\in \lbrace 1, 2 \rbrace$. 
	Then, the received signal of user $j$ is given by
	\begin{align}\label{eq:r}
    	r_j=&~\hat{h}_{0,j}(\sqrt{P_{1}}x_{0,1} + \sqrt{P_{2}}x_{0,2}) + z_{0,j} (\sqrt{P_1}x_{0,1} + \sqrt{P_2}x_{0,2})  \nonumber \\
        &+\hat{h}_{q,j} \sqrt{P_{r}}x_{q,1} + z_{q,j}\sqrt{P_{r}}x_{q,1}+ f_j+n_j
	\end{align}
	where $f_j$ and $n_j$ are the interference from other regions and noise at user $j\in \lbrace 1, 2 \rbrace$, respectively.
	The IoT controller allocates power levels of $P_1$ and $P_2$ to users 1 and 2, respectively, satisfying $P_m=P_1+P_2$.
	The second and fourth terms in \eqref{eq:r} represent the interference caused by the channel estimation errors.
	The interference from the $L$ dominantly interfering regions to user $j$ in the target region is written as 
	\begin{equation}\label{eq:inter-cell-intf}
    	f_j = \sum_{k=1}^{L} \bigg[ \underbrace{h_{0,j}^{(k)} \Big( \sqrt{P_1^{(k)}}x_{0,1}^{(k)}+\sqrt{P_2^{(k)}}x_{0,2}^{(k)}}_{\text{NOMA signals from region $k$}} \Big) + \underbrace{h_{q_k,j}^{(k)}\sqrt{P_r}x_{q_k,1}^{(k)}}_{\text{RRU signal from region $k$}} \bigg], 
	\end{equation}
	where $P_{1}^{(k)}$ and $P_{2}^{(k)}$ are the allocated power levels of the IoT controller for users in region $k$, and $x_{i,j}^{(k)}$ is the symbol transmitted from region $k$. 
	Because the single selection scheme is adopted, only one RRU from each interfering region shares the frequency band with users 1 and 2 of the target region, and its index is denoted as $q_k \in \{1,2,\cdots,S \}$ in region $k$.   
	Assume that $\mathbb{E}[|n_j|^2] = \sigma_n^2$ and $\mathbb{E}[|x_{i,j}|^2] = 1$ for all $i \in \lbrace 0,1,\cdots, S\rbrace$, and $j \in \lbrace1,2\rbrace$. 
	User 1 decodes $x_{0,1}$ and $x_{q,1}$ directly; on the other hand, user 2 performs SIC to cancel $x_{0,1}$ and $x_{q,1}$ from the received signal first, and then decodes $x_{0,2}$. 
	
	This study considers a PS receiver that splits the received signal into two streams for ID and EH with a PS ratio for SWIPT.
	The PS ratio of user $j$ is denoted by $\alpha_j$; therefore, user $j$ uses $\alpha_j$ of the received signal for ID and $1-\alpha_j$ for EH.
	Here, we assume that the interference and noise power cannot be harvested.\footnote{In order to harvest the interference and noise power, they must be in phase with desired signal and large enough that the signal can pass through the rectifier's diode. However, it is very difficult in practical scenarios; therefore, we assume that the interference and noise power cannot be harvested.}
	Then, the signal-to-interference-plus-noise ratio (SINR) for decoding user 1's signal from $r_1$ is obtained as
	\begin{equation}\label{eq:SINR1}
        SINR_1 = \frac{\alpha_1\left(|\hat{h}_{0, 1}|^2P_{1}+|\hat{h}_{q,1}|^2P_{r}\right)}{\alpha_1\left\{|\hat{h}_{0, 1}|^2P_{2} +(L_{0,1}P_m+L_{q,1}P_r)\sigma_{\epsilon}^2\right\}+ \sigma_{f_1}^2 +\sigma_n^2}, 
    \end{equation}
    where $\sigma_{f_j}^2$ is the variance of $f_j$, which can be obtained by taking the averages of each component in \eqref{eq:inter-cell-intf} independently; therefore, $\sigma_{f_j}^2=\sum_{k=1}^L \left(L_{0,j}^{(k)}P_m+L_{q_k,j}^{(k)}P_r\right)$. 
    According to \eqref{eq:SINR1}, the data rate for decoding its own signal for user 1 becomes:
    \begin{equation}\label{eq:Z1}
        Z_1=\text{log}_2 \left( 1 + \frac{\alpha_1 (P_1+v_{1}P_r)}{\alpha_1(P_2+P_{\epsilon,1})+u_1} \right),
    \end{equation}
    where $u_j=(\sigma_{f_j}^2+\sigma_n^2)/{|\hat{h}_{0,j}|^2}$, $v_j={|\hat{h}_{q,j}|^2}/{|\hat{h}_{0,j}|^2}$ and $P_{\epsilon,j}=$ ${(L_{0,j}P_m+L_{q,j}P_r)\sigma_{\epsilon}^2}/{|\hat{h}_{0,j}|^2}$ for $j\in\lbrace 1,2 \rbrace$.
    Similarly, the data rate for decoding the signal of user 1 at user 2 becomes:
	\begin{equation}\label{eq:Z2}
    	Z_2 = \text{log}_2 \left( 1 + \frac{\alpha_2 (P_1+v_{2}P_r)}{\alpha_2(P_2+P_{\epsilon,2})+u_2} \right).
    \end{equation}
	Therefore, the data rate of the signal for user 1 is written as 
    \begin{equation}\label{eq:R1}
        R_1=\text{min}(Z_1, Z_2).
    \end{equation}
    After user 2 performs SIC, the desired symbol of user 2 can be decoded directly. 
    Then, the data rate of user 2 becomes:
    \begin{equation}\label{eq:R2}
    	R_2 = \text{log}_2 \left( 1 + \frac{\alpha_2P_2}{\alpha_2P_{\epsilon,2}+u_2} \right).
    \end{equation}
    In real situations, the RF-to-DC power conversion efficiency function $\eta(\cdot)$ \cite{TMTT2014Jiapin}, which is a nonlinear function of the converter input power, usually an increasing function in the low-input-power region.
    Then, the harvested power at the user $j$ becomes $E_j=\eta(P_{\text{RF},j})P_{\text{RF},j}$,
    where $P_{\text{RF},j}=(1-\alpha_j)\left\{|\hat{h}_{0,j}|^2(P_m+P_{\epsilon,j})+|\hat{h}_{q,j}|^2P_r\right\}$ is the input power of the RF-to-DC rectifier.
    
\section{Max-Sum-Rate Problem}\label{sec:maxsumrate}
    In this section, the sum-rate maximization problem with a minimum rate constraint is studied. 
    Recall that $|\hat{h}_{0,1}|^2\leq|\hat{h}_{0,2}|^2$.
    Let $P_2^{opt}$ be the optimal power allocation for the strong user, and $P_1^{opt}=P_m-P_2^{opt}$ is allocated to the weak user. 
    The sum-rate maximization problem is formulated as
    \begin{align}\label{eq:max-sum problem}
        &\max_{\alpha_1,\alpha_2,P_2} R_{\text{sum}}(\alpha_1,\alpha_2,P_2)\\
        \text{s.t.}~&\min\left\{R_1(\alpha_1,\alpha_2,P_2),R_2(\alpha_2,P_2)\right\}\geq R_{\min} \label{eq:const_min}\\
        &E_1(\alpha_1)\geq\bar{E}_1,~E_2(\alpha_2)\geq\bar{E}_2\label{eq:const_E}\\
        &0\leq\alpha_1\leq1,~0\leq\alpha_2\leq1,~0\leq P_2 \leq P_m, 
    \end{align}
    where $R_{\text{sum}}=R_1+R_2$ and $R_{\min}$ is the minimum data rate constraint.
    Also, $\bar{E}_j$ denotes the minimum harvested energy requirement for user $j$.
    Considering \eqref{eq:Z1}--\eqref{eq:R2}, the constraint \eqref{eq:const_min} can be rewritten with respect to $\alpha_1$ and $\alpha_2$ as follows:
    \begin{align}
        &\alpha_1 \geq  \alpha_{Z_1}= \frac{(2^{R_{\min}}-1)u_1}{P_m+v_1P_r-(2^{R_{\min}}-1)P_{\epsilon,1}-2^{R_{\min}}P_2}\label{eq:alpha_z1_sum}\\
        &\alpha_2 \geq  \alpha_{Z_2}=\frac{(2^{R_{\min}}-1)u_2}{P_m+v_2P_r-(2^{R_{\min}}-1)P_{\epsilon,2}-2^{R_{\min}}P_2}\label{eq:alpha_z2_sum}\\
        &\alpha_2 \geq  \alpha_{R_2}=\frac{(2^{R_{\min}}-1)u_2}{P_2-(2^{R_{\min}}-1)P_{\epsilon,2}}. \label{eq:alpha_r2_sum}
    \end{align}
    The constraints \eqref{eq:const_E} also can be rewritten as 
    \begin{equation}\label{eq:alpha_max}
        \alpha_1 \leq \alpha_{1,\max}~\text{and}~\alpha_2 \leq \alpha_{2,\max},
    \end{equation}
    where $\alpha_{1,\max}$ and $\alpha_{2,\max}$ are the solutions of $E_1(\alpha_1)=\bar{E}_1$ and $E_2(\alpha_2)=\bar{E}_2$, respectively.
    To satisfy \eqref{eq:const_min} and \eqref{eq:const_E}, $P_2$ should satisfy $\alpha_{Z_1}(P_2)\leq\alpha_{1,\max}$, $\alpha_{Z_2}(P_2)\leq\alpha_{2,\max}$, $\alpha_{R_2}(P_2)\leq\alpha_{2,\max}$, $\alpha_{1,\max}\geq0$, and $\alpha_{2,\max}\geq0$. 
    If not, we call this an outage event and $R_{\text{sum}}=0$. 
    We first obtain the optimal PS ratios for given $P_2$ in the following lemma.
    \begin{lemma}\label{lem:opt_alpha}
        For given $P_2$, when $\alpha_{1,\max}>0$ and $\alpha_{2,\max}>0$ are satisfied, the optimal PS ratios that solve the problem \eqref{eq:max-sum problem} are $\alpha_1^*= \alpha_{1,\max}$ and $\alpha_2^*=\alpha_{2,\max}$.
        \begin{IEEEproof}
        For $0\leq\alpha_j\leq\alpha_{j,\max}$, we obtain $\frac{\partial Z_1}{\partial \alpha_1} > 0$, $\frac{\partial Z_2}{\partial \alpha_1} = \frac{\partial R_2}{\partial \alpha_1}=\frac{\partial Z_1}{\partial \alpha_2} = 0$, $\frac{\partial Z_2}{\partial \alpha_2} > 0$, and $\frac{\partial R_2}{\partial \alpha_2} > 0$.
        Because $Z_1$, $Z_2$, and $R_2$ are all non-decreasing functions of both $\alpha_1$ and $\alpha_2$, the sum-rate $R_{\text{sum}}=\min\left\{Z_1,Z_2\right\}+R_2$ is also a non-decreasing function of $\alpha_1$ and $\alpha_2$. Therefore, the optimal PS ratios are the maximum values that can be within ranges.
        \end{IEEEproof}
    \end{lemma}
    We first state Lemma \ref{lem:R_sum_P2}, and the optimal power allocation is obtained using Lemma \ref{lem:opt_alpha} and \ref{lem:R_sum_P2}, as described in Theorem \ref{thm:opt_P2}.
    \begin{lemma}\label{lem:R_sum_P2}
        Define $b_1=P_{\epsilon,1}+{u_1}/{\alpha_1^*}$ and $b_2=P_{\epsilon,2}+{u_2}/{\alpha_2^*}$.
        If $b_1\geq b_2$, $R_{\text{sum}}$ is a non-decreasing function of $P_2$. 
        If $b_1<b_2$, $R_{\text{sum}}$ is a non-increasing function of $P_2$.
        \begin{IEEEproof} 
        By differentiating $Z_1+R_2$ and $Z_2+R_2$ with respect to $P_2$, we have $\frac{\partial (Z_1+R_2)}{\partial P_2} \geq 0$ if $b_1 \geq b_2$, but $\frac{\partial (Z_1+R_2)}{\partial P_2} < 0$ otherwise, and $\frac{\partial (Z_2+R_2)}{\partial P_2}=0$. 
        \end{IEEEproof}
    \end{lemma}
    \begin{theorem}\label{thm:opt_P2}
        The optimal power allocation for maximizing the sum-rate with a minimum rate constraint can be obtained as
        \begin{equation}\label{eq:opt_P2_sum}
            P_2^{*} = \begin{cases}
            \min\left\{P_{Z},P_m\right\},~& b_1\geq b_2 \\
            \max\left\{0,P_{R_2}\right\},~& b_1<b_2 
            \end{cases},
        \end{equation}
        where
        \begin{align}
            &P_Z=\min\left\{P_{Z_1},P_{Z_2}\right\}\\
            &P_{Z_1}=\frac{P_m+v_1P_r-(2^{R_{\min}}-1)b_1}{2^{R_{\min}}}\label{eq:P_z1_sum}\\
            &P_{Z_2}=\frac{P_m+v_2P_r-(2^{R_{\min}}-1)b_2}{2^{R_{\min}}}\label{eq:P_z2_sum}\\
            &P_{R_2}=(2^{R_{\min}}-1)b_2, \label{eq:P_r2_sum}
        \end{align}
        when $P_{R_2}\leq P_Z$ and $[0,P_m]\cap[P_{R_2},P_Z]\neq \phi$. If $P_{R_2}> P_Z$ or $[0,P_m]\cap[P_{R_2},P_Z]= \phi$, an outage event occurs.
        
        \begin{IEEEproof}
        Let the solutions of $\alpha_{Z_1}(P_2)=\alpha_{1,\max}$, $\alpha_{Z_2}(P_2)=\alpha_{2,\max}$, and $\alpha_{R_2}(P_2)=\alpha_{2,\max}$ be $P_{Z_1}$, $P_{Z_2}$, and $P_{R_2}$, respectively. Then, $P_{Z_1}$, $P_{Z_2}$, and $P_{R_2}$ can be obtained as \eqref{eq:P_z1_sum}, \eqref{eq:P_z2_sum} and \eqref{eq:P_r2_sum}, respectively. Because $\alpha_{Z_1}$ and $\alpha_{Z_2}$ are increasing functions of $P_2$, and $\alpha_{R_2}$ is a decreasing function of $P_2$, the constraints \eqref{eq:alpha_z1_sum}--\eqref{eq:alpha_max} can be converted to $P_{R_2}\leq P_2 \leq P_Z=\min\left\{P_{Z_1},P_{Z_2}\right\}$. 
        If $P_{R_2}> P_Z$ or $[0,P_m]\cap[P_{R_2},P_Z]= \phi$, an outage event occurs.
        Otherwise, when $b_1\geq b_2$, $R_{\text{sum}}$ is a non-decreasing function of $P_2$, according to Lemma \ref{lem:R_sum_P2}. Thus, the optimal $P_2^*$ becomes $\min\left\{P_{Z},P_m\right\}$.  
        If $b_1<b_2$, $R_{\text{sum}}$ is a non-increasing for $P_2$, and $P_2^*$ becomes $\max\left\{0,P_{R_2}\right\}$.
        \end{IEEEproof}
    \end{theorem}

\section{Max-Min Fairness Problem}\label{sec:maxminfairness}
    This section provides the optimal power allocation in our framework for the max-min fairness problem, formulated as
    \begin{align}\label{eq:max-min problem}
        &\max_{\alpha_1,\alpha_2,P_2} R_{\text{fair}}(\alpha_1,\alpha_2,P_2)\\
        \text{s.t.}~&Z_2(\alpha_2,P_2)\geq R_{\text{sic}} \label{eq:const_sic}\\
        &E_1(\alpha_1)\geq\bar{E}_1,~E_2(\alpha_2)\geq\bar{E}_2\label{eq:const_E_fair}\\
        &0\leq\alpha_1\leq1,~0\leq\alpha_2\leq1,~0\leq P_2 \leq P_m, 
    \end{align}
    where $R_{\text{fair}}=\min\left\{R_1,R_2\right\}$ and $R_{\text{sic}}$ is the data rate threshold to guarantee the reliability of the SIC operation for user 2. Constraint \eqref{eq:const_sic} can be rewritten with respect to $\alpha_2$ as
    \begin{equation}\label{eq:const_sic2}
        \alpha_2 \geq \alpha_{Z_2}=\frac{(2^{R_{\text{sic}}}-1)u_2}{P_m+v_2P_r-(2^{R_{\text{sic}}}-1)P_{\epsilon,2}-2^{R_{\text{sic}}}P_2}.
    \end{equation}
    The constraints \eqref{eq:const_E_fair} are also rewritten as \eqref{eq:alpha_max}. For given $P_2$, the optimal $\alpha_1$ and $\alpha_2$ are obtained in Lemma \ref{lem:opt_alpha_fair}.
    \begin{lemma}\label{lem:opt_alpha_fair}
        For given $P_2$, if an outage event does not occur, $\alpha_1^*=\alpha_{1,\max}$ and $\alpha_2^*=\alpha_{2,\max}$ are solutions to the problem \eqref{eq:max-min problem}.
        \begin{IEEEproof}
        Because $Z_1$, $Z_2$, and $R_2$ are all non-decreasing functions of both $\alpha_1$ and $\alpha_2$, as explained in the proof of Lemma \ref{lem:opt_alpha}, $R_{\text{fair}}=\min\left\{R_1,R_2\right\}$ is also a non-decreasing function of $\alpha_1$ and $\alpha_2$. Therefore, the optimal $\alpha_1$ and $\alpha_2$ are the maximum values within the ranges that they can have.
        \end{IEEEproof}
    \end{lemma}
    
    We first state Lemma \ref{lem:r1r2_p2}, and Theorem \ref{thm:max-min} provides the optimal power allocation rule that maximizes the user fairness.
    \begin{lemma}\label{lem:r1r2_p2}
        $R_1$ is a decreasing function of $P_2$, and $R_2$ is an increasing function of $P_2$.
        \begin{IEEEproof}
        Differentiating $Z_1$, $Z_2$, and $R_2$ with respect to $P_2$, we have $\frac{\partial Z_1}{\partial P_2} < 0$, $\frac{\partial Z_2}{\partial P_2} < 0$, and $\frac{\partial R_2}{\partial P_2} > 0$.
        Because $Z_1$ and $Z_2$ are decreasing functions of $P_2$, the data rate of user 1, $R_1=\min\left\{Z_1,Z_2\right\}$, is also a decreasing function of $P_2$.
        \end{IEEEproof}
    \end{lemma}
    
    \begin{theorem}
    \label{thm:max-min}
        The optimal power allocation $P_2^{*}$ for the max-min fairness problem \eqref{eq:max-min problem} becomes
        \begin{equation}\label{eq:opt_P2_fair}
            P_2^{*} = \begin{cases}
            P_R,& P_R\leq \min\left\{P_{\text{sic}},P_m\right\},P_{\text{sic}}>0 \\
            \min\left\{P_{\text{sic}},P_m\right\},& P_R>\min\left\{P_{\text{sic}},P_m\right\} ,P_{\text{sic}}>0\\
            \text{outage},&\text{otherwise}
            \end{cases},
        \end{equation}
        where
        \begin{align}
            &P_R=\min\left\{P_{Z_1},P_{Z_2}\right\}\\
            &P_{Z_1}=\frac{-\left(b_1+b_2\right)+\sqrt{\left(b_1+b_2\right)^2+4b_2(P_m+v_1P_r)}}{2}\label{eq:P_z1_fair}\\
            &P_{Z_2}=-b_2+\sqrt{b_2(P_m+v_2P_r+b_2)}\label{eq:P_z2_fair}\\
            &P_{\text{sic}}=\frac{P_m+v_2P_r-(2^{R_\text{sic}}-1)b_2}{2^{R_\text{sic}}}. \label{eq:P_sic}
        \end{align}
        \begin{IEEEproof}
        First, we obtain $P_R$, that is, the solution of the problem \eqref{eq:max-min problem} with the constraint $P_2\geq0$ only.
        For any $P_2\geq 0$, $Z_1(0)>0$, $Z_2(0)>0$, $Z_1(P_m+v_1P_r)=0$ and $Z_2(P_m+v_2P_r)=0$ are satisfied; therefore, we have $R_1(0)>0$ and $R_1\left(P_m+\min\left\{v_1,v_2\right\}P_r\right)=0$.
        Similarly, we also have $R_2(0)=0$ and $R_2\left(P_m+\min\left\{v_1,v_2\right\}P_r\right)>0$. This means that $R_1(P_2)$ and $R_2(P_2)$ should intersect with each other in the interval $P_2\in[0,P_m+\min\left\{v_1,v_2\right\}P_r]$ according to Lemma \ref{lem:r1r2_p2}. 
        Therefore, $R_{\text{fair}}=\min\left\{R_1,R_2\right\}$ is maximized when $R_1=R_2$ for any $P_2\geq0$.
        
        Let $P_{Z_1}$ and $P_{Z_2}$ be the solutions of $Z_1(P_2)=R_2(P_2)$ and $Z_2(P_2)=R_2(P_2)$, respectively. By solving the quadratic equations, the closed-form expressions of $P_{Z_1}$ and $P_{Z_2}$ are obtained as \eqref{eq:P_z1_fair} and \eqref{eq:P_z2_fair}, respectively.
        It can be shown that $P_R=\min\left\{P_{Z_1},P_{Z_2}\right\}$ is the unique solution of $R_1(P_2)=R_2(P_2)$ for $P_2\geq0$.
        
        Now, consider the SIC constraint \eqref{eq:const_sic2}. 
        Let the solution for $\alpha_{Z_2}(P_2)=\alpha_{2,\max}$ be $P_\text{sic}$. Then, $P_\text{sic}$ is obtained as \eqref{eq:P_sic}. Because $\alpha_{Z_2}(P_2)$ is an increasing function of $P_2$, the constraint \eqref{eq:const_sic2} can be converted to $P_2 \leq P_{\text{sic}}$. If $P_\text{sic}<0$, an outage event occurs. 
        Thus, the constraint for the range of $P_2$ is summarized as $0\leq P_2 \leq \min\left\{P_{\text{sic}},P_m\right\}$. 
        When $P_R$ satisfies $0\leq P_R \leq \min\left\{P_{\text{sic}},P_m\right\}$, the optimal power allocation is $P_2^*=P_R$. Otherwise, if $P_R>\min\left\{P_{\text{sic}},P_m\right\}$, $P_2^*=\min\left\{P_{\text{sic}},P_m\right\}$ because $R_{\text{fair}}$ increases with respect to $P_2$ in the interval $[0,P_R]$.
        \end{IEEEproof}
    \end{theorem}

\section{Numerical Results}

This section presents the numerical results of the proposed PS and power allocation rules for the SWIPT-enabled NOMA in DAS. 
Assume that $S=6$, $L=6$, and $K=1$, and the radii of the target region and the local region of the RRU are 20 m and 20/3 m, respectively.
In addition, $\beta=2.5$, $B=5$ MHz, $N_0=-90$ dBm/Hz, $R_{\min}=1$ bps/Hz, $R_{\text{sic}}=0.5$ bps/Hz, and $\bar{E}_1=\bar{E}_2=10~\text{mW}$ are used.
For the RF-to-DC conversion efficiency function $\eta(\cdot)$, we use a 5.8-GHz rectifier conversion efficiency plot \cite{TMTT2014Jiapin}. 
For the power budget $P=P_m+SP_r$, we assume $P_m=10P_r$.
One user is uniformly distributed within the circle with a normalized radius of 0.3, and the other user is uniformly distributed within the ring, which is the region between the inner and outer circles whose normalized radii are 0.8 and 1.0, respectively.
We compare the proposed technique with the following schemes:
\begin{itemize}
\item SWIPT-NOMA: The DAS is not considered. The IoT controller with a power budget of $P$ serves two users using NOMA with appropriate power allocations.
\item SWIPT-OMA in DAS: The IoT controller serves both the strong and weak users but OMA signaling is used. Half of the total bandwidth $B$ is allocated to both the users.
\end{itemize}

\begin{figure}[t]
	\centering
	\includegraphics[width=0.35\textwidth, trim=-0.5cm 0 0 0]{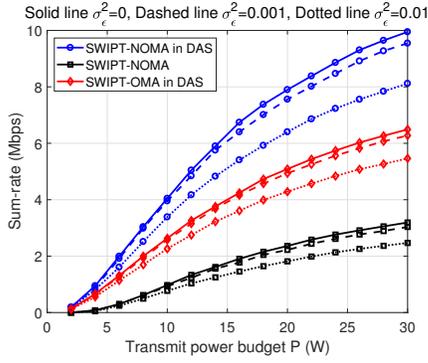}
	\caption{Sum-rate versus transmit power $P$}\label{fig:sumrate_p}
	\vspace{-2mm}
\end{figure}

Fig. \ref{fig:sumrate_p} compares the sum-rate performances. Here, we assume that the sum-rate is zero when $\text{min}(R_1, R_2) < R_{\min}$.
The proposed SWIPT-NOMA in DAS scheme shows much better sum-rate performance than other comparison schemes, even with imperfect CSI.
Overall, schemes using DAS show better performances than SWIPT-NOMA.
This means that DAS can compensate for the deterioration of data rates and energy harvesting caused by long transmission distances.
\begin{figure}[t]
	\centering
	\includegraphics[width=0.35\textwidth, trim=-0.5cm 0 0 0]{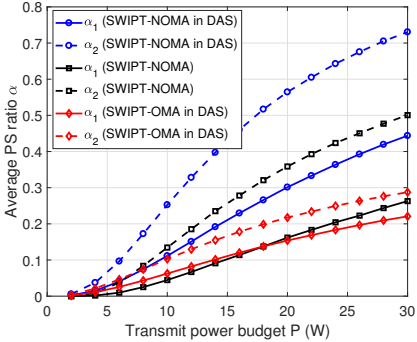}
	\caption{Average $\alpha$ for max-sum-rate versus $P$}\label{fig:alpha_p}
	\vspace{-2mm}
\end{figure}
Fig. \ref{fig:alpha_p} shows the average $\alpha$ for the max-sum-rate problem with $\sigma_\epsilon^2=0.001$.
As $P$ grows, the minimum energy harvesting constraint can be satisfied even with a large $\alpha$; therefore, $\alpha$ also increases.
Because user 1's channel gain is weaker than that of user 2, user 1 generally uses larger portions of the received signal for energy harvesting compared to user 2; therefore, $\alpha_1$ is smaller than $\alpha_2$ in Fig. \ref{fig:alpha_p}.
In addition, the $\alpha$ values of the proposed SWIPT-NOMA in DAS scheme are much larger than those of the comparison schemes, which means that the proposed scheme provides much better energy harvesting efficiency.

\begin{figure}[t]
	\centering
	\includegraphics[width=0.35\textwidth, trim=-0.5cm 0 0 0]{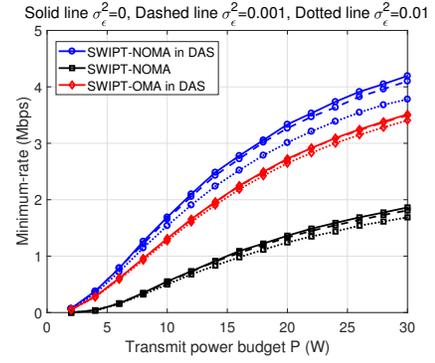}
	\caption{Minimum-rate versus transmit power $P$}\label{fig:minrate_p}
	\vspace{-2mm}
\end{figure}
\begin{figure}[t]
	\centering
	\includegraphics[width=0.35\textwidth, trim=-0.5cm 0 0 0]{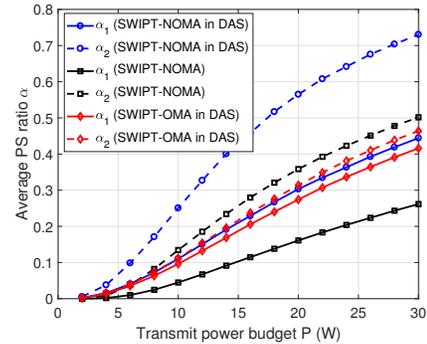}
	\caption{Average $\alpha$ for max-min-fairness versus $P$}\label{fig:alpha_p_min}
	\vspace{-2mm}
\end{figure}

Figs. \ref{fig:minrate_p} and \ref{fig:alpha_p_min} show the minimum rates and the average $\alpha$ with $\sigma_\epsilon^2=0.001$ optimized for the max-min fairness problem, respectively. 
Because there is no minimum rate constraint, the performance degradation according to the imperfect CSI in Fig. \ref{fig:minrate_p} is relatively small compared with Fig. \ref{fig:sumrate_p}.
The performance gap between the proposed scheme and SWIPT-OMA in DAS increases as $P$ grows because the advantage of using NOMA to help the weak user becomes significant with a large $P$.
As in Fig. \ref{fig:alpha_p}, $\alpha_2$ is larger than $\alpha_1$ for all schemes, and the proposed SWIPT-NOMA in DAS scheme has larger $\alpha$ than the others indicating that the energy harvesting of the SWIPT-NOMA in DAS is more efficient than comparison techniques.

\section{Conclusion}
In this paper, we proposed a framework of SWIPT-enabled NOMA in DAS and investigated the synergistic effects of SWIPT, NOMA, and DAS.
In the proposed framework, the optimal PS and power allocation rules for both the max-sum-rate and max-min fairness problems are presented.
Numerical results demonstrate that the sum-rate and user fairness performances of the SWIPT-enabled NOMA can be boosted by using DAS even with imperfect CSI. 

\ifCLASSOPTIONcaptionsoff
  \newpage
\fi

\bibliographystyle{IEEEtran.bst}
\bibliography{IEEEabrv.bib,wcl2021.bib}

\begin{thebibliography}{1}
    \bibitem{CST2015AlFuqaha}
        A. Al-Fuqaha, M. Guizani, M. Mohammadi, M. Aledhari and M. Ayyash, ``Internet of Things: A Survey on Enabling Technologies, Protocols, and Applications,'' \textit{IEEE Commun. Surv. Tut.}, vol. 17, no. 4, pp. 2347-2376, Fourthquarter 2015.

	\bibitem{ISIT2008Varshney}
		L. R. Varshney, ``Transporting information and energy simultaneously,'' in \textit{Proc. IEEE Int. Symp. Info. Theory}, Jul. 2008, pp. 1612--1616.
		
	\bibitem{CommMag2014Krikidis}
	    I. Krikidis, S. Timotheou, S. Nikolaou, G. Zheng, D. W. K. Ng and R. Schober, ``Simultaneous wireless information and power transfer in modern communication systems,'' \textit{IEEE Commun. Mag.}, vol. 52, no. 11, pp. 104--110, Nov. 2014.

	\bibitem{ComMag2017Ding}
		Z. Ding, \textit{et al.}, ``Application of non-orthogonal multiple access in LTE and 5G networks'' \textit{IEEE Commun. Mag.}, vol. 55, pp. 185--191, Feb. 2017.
		
	\bibitem{JSAC2016Liu}
    	Y. Liu, Z. Ding, M. Elkashlan, and H. V. Poor, ``Cooperative non-orthogonal multiple access with simultaneous wireless information and power transfer,'' \textit{IEEE J. Sel. Areas Commun.}, vol. 34, no. 4, pp. 938--953, Apr. 2016.
		
	\bibitem{JSAC2019Tang}
    	 J. Tang et al., ``Energy efficiency optimization for NOMA with SWIPT,'' \textit{IEEE J. Sel. Areas Commun.}, vol. 13, no. 3, pp. 452--466, Jun. 2019.
    
    \bibitem{TWC2007Choi}
		W. Choi and J. G. Andrews, ``Downlink performance and capacity of distributed antenna systems in a multicell environment,'' \textit{IEEE Trans. Wireless Commun.}, vol. 6, no. 1, pp. 69--73, Jan. 2007.
	
	\bibitem{TCOM2015Kim}
		H. Kim, S. Lee, C. Song, K. Lee and I. Lee, ``Optimal Power Allocation Scheme for Energy Efficiency Maximization in Distributed Antenna Systems," \textit{IEEE Trans. Commun.}, vol. 63, no. 2, pp. 431--440, Feb. 2015.
	    
	\bibitem{IoT2018Huang}
    	Y. Huang, M. Liu and Y. Liu, ``Energy-Efficient SWIPT in IoT Distributed Antenna Systems,'' \textit{IEEE Internet Things J.}, vol. 5, no. 4, 2018, pp. 2646--2656
    	
	\bibitem{Access2021Kim}
	    D. Kim and M. Choi, ``Non-Orthogonal Multiple Access in Distributed Antenna Systems for Max-Min Fairness and Max-Sum-Rate,'' \textit{IEEE Access}, vol. 9, pp. 69467--69480, 2021.

	\bibitem{3GPP}
		“Study on downlink multiuser superposition transmission for LTE,” TSG RAN Meeting 67, 3rd Generation Partnership Project (3GPP), Tech. Rep. RP-150496, Mar. 2015.
		
	\bibitem{VTC2019Zamani}
		M. R. Zamani, M. Eslami, M. Khorramizadeh, and Z. Ding, ``Energy efficient power allocation for NOMA with imperfect CSI,'' \textit{IEEE Trans. Veh. Tech.}, vol. 68, no. 1, pp. 1009--1013, Jan. 2019.

	\bibitem{TMTT2014Jiapin}
	    G. Jiapin, Z. Hongxian, and Z. Xinen, ``Theoretical analysis of RF-DC conversion efficiency for class-F rectifiers,'' \textit{IEEE Trans. Microw. Theory Tech.}, vol. 62, no. 4, pt. 2, pp. 977–985, Apr. 2014.
	    
\end{thebibliography}

\end{document}